\documentclass[aps,reprint,twocolumn,showpacs,superscriptaddress]{revtex4-1}
\usepackage{graphicx}
\usepackage{xcolor}
\usepackage{epstopdf}
\usepackage{dcolumn}
\usepackage{bm}
\usepackage{color}
\usepackage{subfigure}
\usepackage{amsmath}

\begin{document}


\title{Non-Hermitian nodal-line semimetals with an anomalous bulk-boundary correspondence}

\author{Huaiqiang Wang}
\affiliation{National Laboratory of Solid State Microstructures and School of Physics, Nanjing University, Nanjing 210093, China}

\author{Jiawei Ruan}
\affiliation{National Laboratory of Solid State Microstructures and School of Physics, Nanjing University, Nanjing 210093, China}

\author{Haijun Zhang}
\email{zhanghj@nju.edu.cn}
\affiliation{National Laboratory of Solid State Microstructures and School of Physics, Nanjing University, Nanjing 210093, China}
\affiliation{Collaborative Innovation Center of Advanced Microstructures, Nanjing University, Nanjing 210093, China}






\begin{abstract}
Recently, topological quantum states of non-Hermitian systems, exhibiting rich new exotic states, have attracted great attention in condensed-matter physics. As for the demonstration, most of non-Hermitian topological phenomena previously focused on are in one- and two-dimensional systems. Here, we investigate three-dimensional non-Hermitian nodal-line semimetals in the presence of a particle gain-and-loss perturbation. It is found that this perturbation will split the original nodal ring into two exceptional rings (ERs). The topological nature of the bulk electronic structure is characterized by two different topological invariants, namely, the vorticity and the winding number defined for a one-dimensional loop in momentum space, both of which are shown to take half-integer (integer) values when an odd (even) number of ERs thread through the loop. The conventional bulk-surface correspondence in non-Hermitian nodal-line semimetals is found to break down, where the surface zero-energy flat bands are no longer bounded by projections of bulk ERs. Alternatively, a macroscopic fraction of the bulk eigenstates can be localized near the surface, thus leading to the so-called non-Hermitian skin effect.
\end{abstract}


\maketitle


\section{Introduction}
The studies on topological states of Hermitian systems, including topological insulators~\cite{Hasan2010,Qi2010,chiu2016}, topological superconductors~\cite{Qi2010,Fu2008,Qi2010,Sau2010,he2017chiral}, and topological semimetals~\cite{Armitage2018,Yang2014, gao2016classification,yan2017topological,Wan2011,Xu2011, Lu2012,Weng2015,Xu2015a,Lv2015a,Ruan2016a,Ruan2016b,liu2014discovery, bradlyn2016beyond,wang2016hourglass,Yan2017Nodal, Chen2017Topological,Bomantara2016,Wang2017Line}, have profoundly deepened our understandings of symmetries and topology in condensed-matter physics. Topological states can be characterized by corresponding topological invariants defined from bulk band structures, which ensure the existence of gapless boundary states through the celebrated bulk-boundary correspondence. Among various topological materials, nodal-line semimetals have attracted much interest and have been intensively studied both theoretically~\cite{Kim2015Dirac,Bian2016Drumhead,Yu2015Topological,Burkov2011,Phillips2014,Mullen2015,Yan2016,Fang2015,Zhang2016Quantum, Hirayama2017Topological} and experimentally~\cite{Bian2016Topological,Wu2016Dirac,Schoop2016Dirac}. They have band degeneracies along lines in momentum space, and possess drumhead-like surface states, which hold a potential possibility for realizations of surface superconductivity and surface magnetism when electron-electron correlation is introduced~\cite{Kopnin2011,Roy2017}.

Very recently, there has been growing interest in topological states of non-Hermitian systems~\cite{Gong2018}. Non-Hermiticity is ubiquitous in a diverse range of situations, including open quantum systems~\cite{Carmichael1993, Rudner2009, Choi2010, Diehl2011Topology, Lee2014, Malzard2015, Zhen2015Spawning}, optical systems with gain and loss~\cite{Klaiman2008, Regensburger2012Parity, Bittner2012, Hodaei2014Parity, Feng2014Single, Elganainy2018Non, Zhou2018Observation, Takata2018Photonic}, and interacting/disordered systems~\cite{Kozii2017Non, Papaj2018Bulk, Zyuzin2018, Shen2018Quantum, Zhao2018Condition}. The interplay between non-Hermiticity and topology leads to quite distinct properties in non-Hermitian systems, such as the breakdown of the conventional bulk-boundary correspondence~\cite{Xiong2017Why, Lee2016, Yao2018Edge, Yao2018Non, Kawabata2018,  Alvarez2018Topological, Kunst2018, jin2018bulk}, the emergence of anomalous edge states~\cite{Lee2016, Yao2018Edge, Yao2018Non}, and the anomalous localization of bulk eigenstates (``non-Hermitian skin effect'')~\cite{Yao2018Edge, Yao2018Non, Martinez2018, lee2018anatomy}. It has also been shown that non-Hermitian topology could manifest itself in some interesting transport phenomena~\cite{Rudner2016Survival, Ostahie2016, Hu2017, Zhang2017Trans, Avila2018, Yang2018, Mcdonald2018Phase, Harari2018Topological,  Chen2018Hall, Philip2018Loss, Longhi2015Non}, for example, the deviation of the Hall conductance of the edge state from the quantized Chern number~\cite{Philip2018Loss, Chen2018Hall}, one-way transport in low-dimensional lattices by an imaginary gauge field~\cite{Longhi2015Non}, and the topological insulator laser~\cite{Harari2018Topological}.

Up to now, most non-Hermitian topological phenomena previously studied are limited in one-dimensional (1D) and two-dimensional (2D) systems~\cite{Lee2016, Yao2018Edge, Yao2018Non, Kawabata2018,  Alvarez2018Topological, Kunst2018, Martinez2018, Leykam2017, Shen2018, Lieu2018, Klett2017, Hu2011, Esaki2011, Zhou2018Non}, and much less effort has been devoted to three-dimensional (3D) systems~\cite{Xu2017Weyl, Gonz2017, Cerjan2018, Carlstrom2018, Yang2018Nodal}. In this work, we investigate both continuum and lattice models of non-Hermitian nodal-line semimetals in the presence of a particle gain-and-loss term. It is found that such a non-Hermitian perturbation will split each nodal ring into two exceptional rings (ERs). With increasing strength of this perturbation, some of the ERs may shrink and eventually vanish. To characterize the topological property of the bulk band structure, two different topological invariants are used: (1) One is the vorticity~\cite{Shen2018} of a loop around the exceptional points (EPs) generated by cutting the ERs with a 2D slice in the cylinder coordinate. (2) The other is the winding number for a loop in the 3D momentum space, which stems from the chiral symmetry and can be calculated through the definition of a complex angle~\cite{Leykam2017, Yin2018}. Both invariants take fractional (integer) values when the loop is threaded by an odd (even) number of ERs. Under open boundary conditions (OBCs), the drumhead-like surface bands are no longer  bounded by the projections of bulk ERs, thus suggesting the breakdown of conventional bulk-surface correspondence in Hermitian nodal-line semimetals. Intriguingly, not only the drumhead-like surface bands but also a macroscopic fraction of bulk states are found to be localized on the surface, which could be explained by dimensional reduction to 1D non-Hermitian lattice models.

This paper is organized as follows. In Sec. II, we first study the bulk-band structure of non-Hermitian nodal-line semimetals through a simple continuum model in Sec. II A and then introduce the two topological invariants, namely, the vorticity and the winding number, in Secs. II B and II C, respectively, to characterize the bulk topology. In Sec. III, we address the issue of non-Hermitian bulk-boundary correspondence, where a lattice model is used to illustrate the band structures under periodic boundary conditions (PBCs) and OBCs in Sec. III A. The skin effect of non-Hermitian nodal-line semimetals is discussed in Sec. III B. Section IV concludes this paper.
\section{Bulk band from the continuum model}
\subsection{Model description}
\begin{figure}[t]
\includegraphics[clip,angle=0,width=8cm]{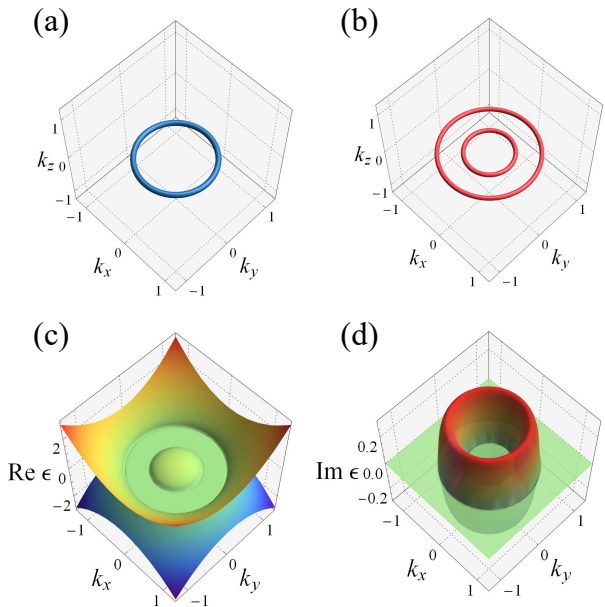}
\caption{Illustration of the nodal rings in the $k_{z}=0$ plane in the (a) absence and (b) presence of the non-Hermitian term $i\gamma_{z}\tau_{z}$.  (c) The real part and (d) the imaginary part of the energy dispersion in the $k_{z}=0$ plane with the same parameters as in (b). The parameters are chosen as $m=0.5$, $B=v_{z}=1$, and $\gamma_{z}=0.3$ for the continuum model, and remain unchanged in the following unless otherwise specified.}
\end{figure}
A typical two-band spinless nodal-line semimetal can be described by the simple continuum model Hamiltonian~\cite{Fang2015, Yan2016}:
\begin{equation}
\label{Hermitian H}
H(\mathbf{k})=\epsilon_{0}(\mathbf{k})\tau_{0}+(m-Bk^{2})\tau_{x}+v_{z}k_{z}\tau_{z},
\end{equation}
where $k^{2}=k_{x}^{2}+k_{y}^{2}+k_{z}^{2}$, $\tau_{i}$ ($i=x,y,z$) are Pauli matrices acting in the two-orbital subspace, $\tau_{0}$ is the identity matrix, $v_{z}$ denotes the Fermi velocity along the $k_{z}$ direction, and $m$ and $B$ are parameters with the dimension of energy and inverse energy, respectively~\cite{Yan2016}. When $mB>0$, the conduction and valence bands touch along the nodal ring located in the $k_{z}=0$ plane at $k_{x}^{2}+k_{y}^{2}=m/B$ [see Fig. 1(a)], while for $mB<0$, the system lies in the trivial insulator phase with an energy gap. Without loss of generality and for simplicity, henceforth, unless stated explicitly, $m$, $B$, and $v_{z}$ are assumed to be positive. The Hermitian nodal ring is protected by the combined inversion and time-reversal symmetry $PT$~\cite{Zhang2016Quantum}, which can be simply represented as the complex conjugate $K$ in a proper orbital basis. Such a symmetry imposes a reality condition on the Hamiltonian as $H(\mathbf{k})=H(\mathbf{k})^{*}$ and restricts the $\tau_{y}$ term to zero. This reduces the number of equations for band degeneracies to two, thus ensuring the emergence of line nodes in the 3D momentum space. In addition, when $\epsilon_{0}(\mathbf{k})=0$, the Hamiltonian in Eq. (\ref{Hermitian H}) also satisfies the chiral symmetry, $\tau_{y}H(\mathbf{k})\tau_{y}=-H(\mathbf{k})$, which constrains the whole nodal ring to zero energy.

In the presence of a non-Hermitian term $i \gamma_{z}\tau_{z} (\gamma_{z}>0)$ associated with particle gain and loss for the two orbitals,  the Hamiltonian becomes:
\begin{equation}
\label{Hamiltonian}
H(\mathbf{k})=\epsilon_{0}(\mathbf{k})\tau_{0}+(m-Bk^{2})\tau_{x}+(v_{z}k_{z}+i\gamma_{z})\tau_{z}.
\end{equation}
The energy is now obtained as
\begin{equation}
\label{energy}
E_{\pm}=\epsilon_{0}(\mathbf{k})\pm\sqrt{(m-Bk^{2})^{2}+v_{z}^{2}k_{z}^{2}-\gamma_{z}^{2}+2iv_{z}k_{z}\gamma_{z}},
\end{equation}
which is generally complex for nonzero $\gamma_{z}$. Since $\epsilon_{0}(\mathbf{k})$ has no effect on band crossings and eigenstates, unless otherwise specified, it will be set to zero henceforth. Note that the non-Hermitian $i \gamma_{z}\tau_{z} (\gamma_{z}>0)$ term explicitly breaks the $PT$ symmetry of the Hermitian model but preserves the chiral symmetry in the absence of the constant-energy term. To see the fate of the original nodal ring, we focus on the $k_{z}=0$ plane, where the energy becomes $E_{\pm}=\pm\sqrt{\big(m-Bk_{\parallel}^{2}\big)^{2}-\gamma_{z}^{2}}$, with $k_{\parallel}\equiv\sqrt{k_{x}^{2}+k_{y}^{2}}$. When $\gamma_{z}<m$, the original nodal ring splits into two ERs characterized by $Bk^{2}_{\parallel}=m\pm\gamma_{z}$, as shown in Fig. 1(b). In the $k_{z}=0$ plane, the energy is purely real both inside the inner ER and outside the outer ER, while it is purely imaginary between the two ERs, as demonstrated in Figs. 1(c) and 1(d), respectively. With increasing $\gamma_{z}$, the inner ER shrinks and vanishes beyond the critical value of $\gamma_{z}=m$, where it becomes a point. Intriguingly, an ER appears even for the original gapped phase with negative $m$, as long as $\gamma_{z}>|m|$ is satisfied.

Before further discussion, several points need to be clarified concerning the non-Hermitian perturbations and corresponding band degeneracies. First, generally speaking, in non-Hermitian systems, the number of conditions for two-band crossings is two instead of three in the Hermitian case~\cite{Berry2004}, and therefore 1D nodal lines are realizable in 3D non-Hermitian systems with three tunable momentum parameters even in the absence of any symmetry, as is the case with the Weyl ER~\cite{Xu2017Weyl} and the present model in Eq. (\ref{Hamiltonian}) regardless of the $\epsilon_{0}(\mathbf{k})$ term. Second, if we consider a $PT$-symmetric non-Hermitian perturbation such as an $i\gamma_{y}\tau_{y}$ term to the Hermitian Hamiltonian in Eq. (\ref{Hermitian H}), the nodal ring may even evolve into an exceptional surface~\cite{budich2018symmetry, okugawa2018topological, zhou2018exceptional}. Third, in contrast to two previous papers, namely, Refs.~\cite{Carlstrom2018} and~\cite{Yang2018Nodal}, both of which mainly investigate the possibility of realizing exceptional links from nodal-line semimetals under certain non-Hermitian perturbations, in this paper, based on the nodal-line semimetals under a simple gain-and-loss perturbation, we focus on the topological properties of ordinary ERs without links, as well as the anomalous bulk-surface correspondence.

\begin{figure}[t]
	\includegraphics[clip,angle=0,width=8.5cm]{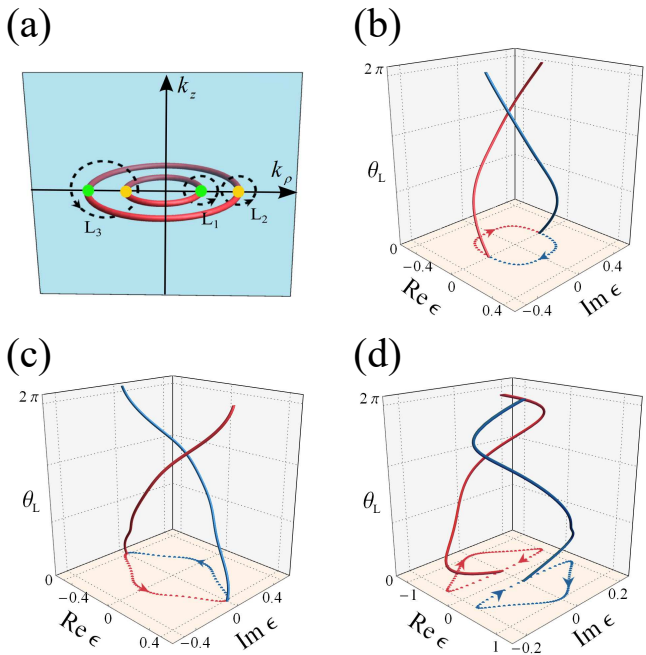}
	\caption{ (a) Schematic view of EPs in the $k_{\rho}$-$k_{z}$ plane. Here, green and yellow colors represent the $1/2$ and $-1/2$ vorticities, respectively. Three dashed loops are marked as $L_{1,2,3}$ for the evolution of the complex eigenvalues. The evolution of the two complex eigenvalues of EPs along loops (b) $L_{1}$, (c) $L_{2}$, and (d) $L_{3}$, which are parameterized by $\theta_{L}\in[0,2\pi)$. Their projections onto the complex plane are also presented.}
\end{figure}

\subsection{The vorticity}
In contrast to Hermitian band degeneracies consisting of distinct eigenvectors, EPs are ubiquitous in non-Hermitian band structures, where not only the eigenvalues but also the eigenvectors coalesce with each other, thus rendering the corresponding Hamiltonian defective and nondiagonizable~\cite{Berry2004}. When encircling an EP, the constitutive bands get exchanged due to the square root taken in Eq. (\ref{energy}), and two loops are required to return to the initial state~\cite{Berry2004, Moiseyev, Heiss2012, Mailybaev2005, Dembowski2004, kim2013braid, Lee2016, Leykam2017, Shen2018}. In order to characterize the ERs, we adopt the cylinder-like coordinate and divide each ER into a collection of EPs residing in the 2D $k_{\rho}$-$k_{z}$ slice [see Fig. 2(a); here, $k_{\rho}$ is allowed to take negative values, which should be distinguished from the conventional cylinder coordinate]. After this decomposition, we can then resort to the concept of vorticity \cite{Shen2018} to characterize each EP.

First, we consider the case with both the inner and outer ERs ($\gamma_{z}<m$), which are located in the $k_{z}=0$ plane at $k_{\parallel}=\sqrt{(m-\gamma_{z})/B}$ and $\sqrt{(m+\gamma_{z})/B}$, respectively. For each $k_{\rho}$-$k_{z}$ slice, altogether four EPs appear at $(k_{\rho},k_{z})=(k_{\pm}^{s},0)$, as shown in Fig. 2(a), with
\begin{equation}
\begin{split}
\label{EP}
k_{\pm}^{s}=\pm\sqrt{(m-s\gamma_{z})/B},
\end{split}
\end{equation}
where $s=+1$ $(-1)$ for the EPs from the inner (outer) ER. In fact, these EPs can be understood from the non-Hermitian-term-induced splittings of the original Dirac points at $(\pm\sqrt{m/B},0)$ in the 2D $k_{\rho}$-$k_{z}$ slice. By expanding the low-energy effective Hamiltonian to linear order around each EP, we obtain
\begin{equation}
H_{\pm}^{s}(\mathbf{q})=(s\gamma_{z}-2Bk_{\pm}^{s}q_{\rho})\tau_{x}+(v_{z}q_{z}+i\gamma_{z})\tau_{z}.
\end{equation}
The dispersion to the leading order of $\mathbf{q}$ is then derived as
\begin{equation}
\label{Dispersion}
E_{\pm,\lambda}^{s}(\mathbf{q})=\lambda\sqrt{2\gamma_{z}(-sv_{\pm}^{s}q_{\rho}+iv_{z}q_{z})},
\end{equation}
where $v_{\pm}^{s}=2Bk_{\pm}^{s}$ and $\lambda=\pm1$ for the two branches of bands. Following Ref. \cite{Shen2018}, the vorticity of each EP can be calculated as
\begin{equation}
\nu_{\pm}^{s}=-\frac{1}{2\pi}\oint_{\Gamma}\nabla_{\mathbf{q}}\mathrm{arg}[E_{\pm,+}^{s}(\mathbf{q})-E_{\pm,-}^{s}(\mathbf{q})]\cdot d\mathbf{q}=\pm\frac{s}{2},
\end{equation}
where $\Gamma$ is a closed loop encircling the EP. A nonzero vorticity for such a contractible closed loop in momentum space indicates a band degeneracy surrounded by $\Gamma$~\cite{Shen2018}. It should be emphasized that the fractional vorticity is an inherent property of the EP unique to non-Hermitian systems and is well defined in the absence of any symmetry.

As an illustration, in Figs. 2(b) and 2(c), we numerically plot the evolution paths of the two bands along the loops $L_{1}$ and $L_{2}$ around the inner and outer EPs $k_{+}^{+}$ and $k_{+}^{-}$, respectively, both of which are parameterized by $\theta_{L}\in[0,2\pi]$. It can be seen that around both $k_{+}^{+}$ and $k_{+}^{-}$, the two bands get switched at $\theta_{L}=2\pi$.  However, they wind around each other in opposite directions, namely, clockwise (counterclockwise) for $k_{+}^{+}$ ($k_{+}^{-}$) with $v_{+}^{+}=1/2$ ($v_{+}^{-}=-1/2$), as clearly demonstrated by their projections to the complex plane. More generally, when the loop encloses an odd number of EPs, the two bands swap with each other and the vorticity takes a half-integer value, while when an even number of EPs is enclosed, the vorticity becomes an integer, and the two bands return to their original states, as exemplified by the loop $L_{3}$ enclosing both $k_{-}^{+}$ and $k_{-}^{-}$ in Fig. 2(a), with the bands' evolution shown in Fig. 2(d).

In addition, with increasing $\gamma_{z}$, the two EPs from the inner ER approach each other until $\gamma_{z}=m$, where they meet and annihilate as a result of their opposite vorticities~\cite{Shen2018}, which accounts for the disappearance of the inner ER when $\gamma_{z}>m$.

\subsection{The winding number}
To fully capture the topological property of the bulk band, the above calculation of vorticity is insufficient. For example, it cannot distinguish between a loop enclosing two EPs with opposite vorticities and a loop enclosing no EPs since both loops exhibit zero vorticity. Moreover, as the vorticity depends only on the energies, it fails to provide topological properties concerning the eigenstates such as the Berry phase~\cite{Mailybaev2005, Lieu2018}. However, in non-Hermitian systems, when encircling an EP, two loops are needed to return to the original state, thus making it problematic to calculate the conventional Berry phase for a single loop. To circumvent this, in this section, we will calculate the winding number originating from the chiral symmetry of the non-Hermitian Hamiltonian in Eq. (\ref{Hamiltonian}) in the absence of $\epsilon_{0}(\mathbf{k})$, which has been shown to be closely related to the non-Hermitian generalization of the Berry phase~\cite{Leykam2017, garrison1988complex}. By treating $k_{x}$ and $k_{y}$ as parameters, the winding number can be defined for every 1D chain along the $k_{z}$ direction as \cite{Leykam2017, Yin2018, Zhou2017Dynamical}
\begin{equation}
\label{winding number}
w=\frac{1}{2\pi}\int_{-\infty}^{\infty}dk_{z}\partial_{k_{z}}\phi,
\end{equation}
where $\phi\equiv\arctan(h_{x}/h_{z})=\arctan[(m-Bk^{2})/(v_{z}k_{z}+i\gamma_{z})]$, with $h_{x}$ and $h_{z}$ representing the components of the $\tau_{x}$ and $\tau_{z}$ terms, respectively, in $H$. (If the alternative definition $\phi\equiv\arctan(h_{z}/h_{x})$ is used, the final result of the winding number will differ only by a sign reversal.) Note that the presence of the non-Hermitian term indicates that $\phi$ is generically complex.

\begin{figure*}[t]
\includegraphics[clip,angle=0,width=13cm]{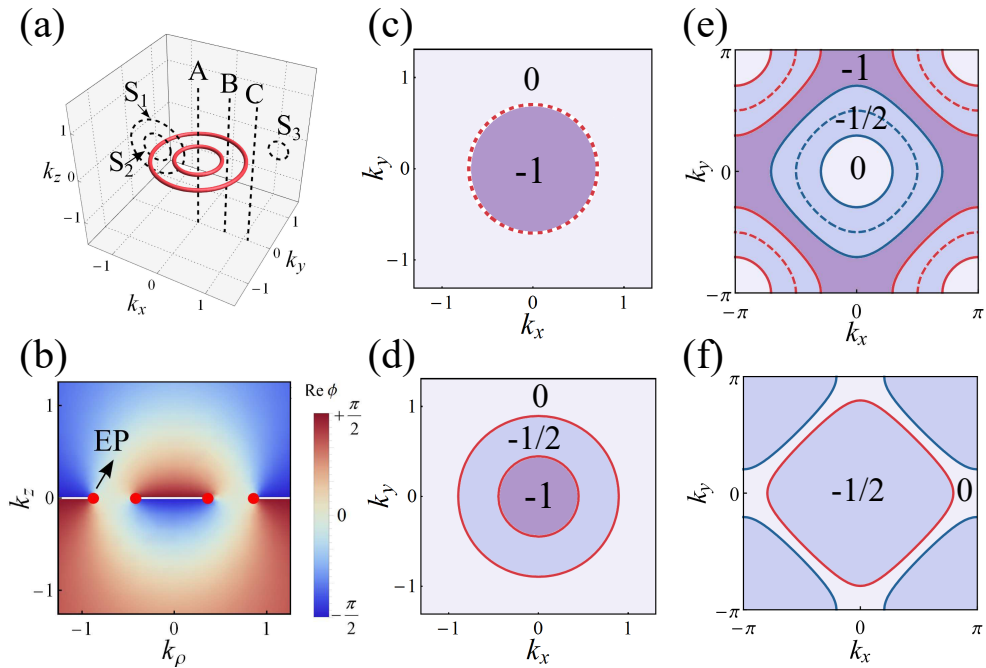}
\caption{(a) Representative 1D loops in the momentum space are plotted as dashed lines (e.g., S$_{1,2,3}$, A, B, C) for the calculation of the winding number. (b) The distribution of Re($\phi$) in the 2D $k_{\rho}$-$k_{z}$ slice, where the red dots represent EPs. The winding number as a function of $k_{x}$ and $k_{y}$ in the (c) absence and (d) presence of the $i\gamma_{z}\tau_{z}$ term ($\gamma_{z}=0.3$). The nodal rings in the $k_{z}=0$ (red lines) and $k_{z}=\pi$ (blue lines) planes for the lattice model with (e) $\gamma_{z}=0.6$ and (f) $\gamma_{z}=1.2$, respectively. The corresponding winding number as a function of $k_{x}$ and $k_{y}$ is also presented. The parameters $m=3$, $B=0.5$, $v_{z}=1$ are taken in (e) and (f).}
\end{figure*}

When $m>\gamma_{z}$, two ERs appear in the $k_{z}=0$ plane, as shown before. Considering the rotational symmetry of the system, we numerically present the real part of $\phi$ for a 2D $k_{\rho}$-$k_{z}$ slice in Fig. 3(b), with the parameters $m=0.5$, $B=v_{z}=1$, and $\gamma_{z}=0.3$. Here, Re($\phi$) is an odd function of $k_{z}$ and at $k_{z}=0$, it is continuous when $k_{\rho}$ lies between the two ERs [line B in Fig. 3(a)], while for $k_{\rho}$ outside this range [lines A and C in Fig. 3(a)], it is discontinuous with a $\pi$ jump. Nevertheless, the real part of $\partial_{k_{z}}\phi$ is always continuous with no such jumps, thus validating Eq. (\ref{winding number}). In contrast, the imaginary part of $\phi$ is found to be an even and continuous function of $k_{z}$ here, suggesting its derivative Im($\partial_{k_{z}}\phi$) is an odd function and $\mathrm{Im}\phi(k_{z}\rightarrow\infty)=\mathrm{Im}\phi(k_{z}\rightarrow-\infty)$, which consequently does not contribute to the integral in Eq. (\ref{winding number}). Finally, the winding number can be explicitly derived as (see Appendix A for the detailed derivation)
\begin{equation}
\label{w result}
w=\left\{
    \begin{array}{ll}
      -1, & \hbox{for $|k_{\rho}|<k_{\mathrm{in}}$,} \\
      -\frac{1}{2}, & \hbox{for $k_{\mathrm{in}}<|k_{\rho}|<k_{\mathrm{out}}$,} \\
      0, & \hbox{for $|k_{\rho}|>k_{\mathrm{out}}$,}
    \end{array}
  \right.
\end{equation}
where $k_{\mathrm{in}}=\sqrt{(m-\gamma_{z})/B}$ and $k_{\mathrm{out}}=\sqrt{(m+\gamma_{z})/B}$, which are the radii of the inner and outer ERs, respectively. This result is numerically supported by the phase diagram of $w$ as a function of both $k_{x}$ and $k_{y}$ in Fig. 3(d), where the boundaries between regions of different $w$ values (solid red lines) exactly correspond to locations of the bulk ER. As a comparison, in Fig. 3(c), we present the $w$ phase diagram with the same parameters in the absence of the $\gamma_{z}$ term. The merging of the two ERs into the Hermitian nodal ring leads to the disappearance of the region with a fractional value of $w$, and recovers the result for a Hermitian nodal-line semimetal.

The emergence of the fractional value $w=-1/2$ and integer value $w=-1$ can be understood as follows. Although the values of $\phi$ are found to differ by $\pi$ for the two opposite limits $k_{z}\rightarrow\infty$ and $k_{z}\rightarrow-\infty$, their derivatives $\partial_{k_{z}}\phi$ turn out to be the same, thus enabling us to reasonably compact the integral line into a loop by connecting $k_{z}=\infty$ to $k_{z}=-\infty$ (the compactness will be quite natural for a Bloch Hamiltonian in a lattice model with PBCs). As a result, lines A, B, and C, are topologically equivalent to loops $S_{1}$, $S_{2}$, and $S_{3}$, respectively, which are threaded by two, one, and zero ERs. Since the $S_{1}$ loop encloses two EPs, the winding number can be proved to be $\pm1$~\cite{Lee2016, Yin2018}, with the non-Hermitian-generalized Berry phase $\phi_{B}=\pi$ (mod $2\pi$). This can be understood from the $\pi$ Berry phase for a loop encircling the unperturbed Hermitian nodal ring. For the $S_{2}$ loop encircling only one EP, the winding number is found to take fractional values $\pm1/2$~\cite{Lee2016}, which is related to the fact that $\phi_{B}=\pi$ (mod $2\pi$) only after a path circles twice around an EP~\cite{Lee2016, Mailybaev2005, Dembowski2004}. For the $S_{3}$ loop enclosing no EPs, the winding number should obviously take the trivial value zero with $\phi_{B}=0$ (mod $2\pi$). Consequently, the winding number is related to the non-Hermitian Berry phase as
\begin{equation}
 w\pi\equiv\phi_{B} (\mathrm{mod}\ 2\pi).
\end{equation}
Note that the $\pm\frac{\pi}{2}$ phase here means the ``averaged'' phase for a loop~\cite{Mailybaev2005}.

When $m<\gamma_{z}$, only the outer ER remains, and it is evident from the above analysis that $w=-1/2$ ($w=0$) inside (outside) this ER.

In the above discussion, the constant energy term $\epsilon_{0}(\mathbf{k})$ has been neglected to satisfy the chiral symmetry. However, although the presence of such a term explicitly breaks the chiral symmetry and invalidates the definition of the winding number, the Berry phase argument remains the same since the $\epsilon_{0}(\mathbf{k})$ term does not change the eigenstates.
\section{Anomalous bulk-surface correspondence}
In Hermitian systems, by virtue of bulk-boundary correspondence, the emergence of topological surface (edge) states is ensured by relevant topological invariants of bulk bands under PBCs. This rule holds true for Hermitian nodal-line semimetals, where drumhead surface states (flat bands) are expected to be bounded by the projections of bulk nodal rings onto the surface Brillouin zone (BZ)~\cite{Yu2015Topological, Kim2015Dirac, Bian2016Drumhead, Wang2017Line}. However, the generalization of such correspondence to non-Hermitian systems is problematic and has been shown to break down in certain systems~\cite{Lee2016, Xiong2017Why, Yao2018Edge, Yao2018Non, Kawabata2018, Kunst2018, Alvarez2018Topological, jin2018bulk, Martinez2018}, such as the non-Hermitian Su-Schrieffer-Heeger (SSH) model~\cite{Yao2018Edge, Kunst2018} and the non-Hermitian Chern insulator~\cite{Yao2018Non, Kunst2018, Kawabata2018}. Intriguingly, under OBCs, even a macroscopic number of bulk eigenstates become localized near the boundary, producing the so-called non-Hermitian skin effect~\cite{Xiong2017Why, Yao2018Edge, Yao2018Non, Martinez2018, Kunst2018}. In this section, we will inspect such effects for a lattice model of a non-Hermitian nodal-line semimetal under PBCs and OBCs.

\subsection{Bloch band from lattice model}
\begin{figure*}[htp]
\includegraphics[clip,angle=0,width=17cm]{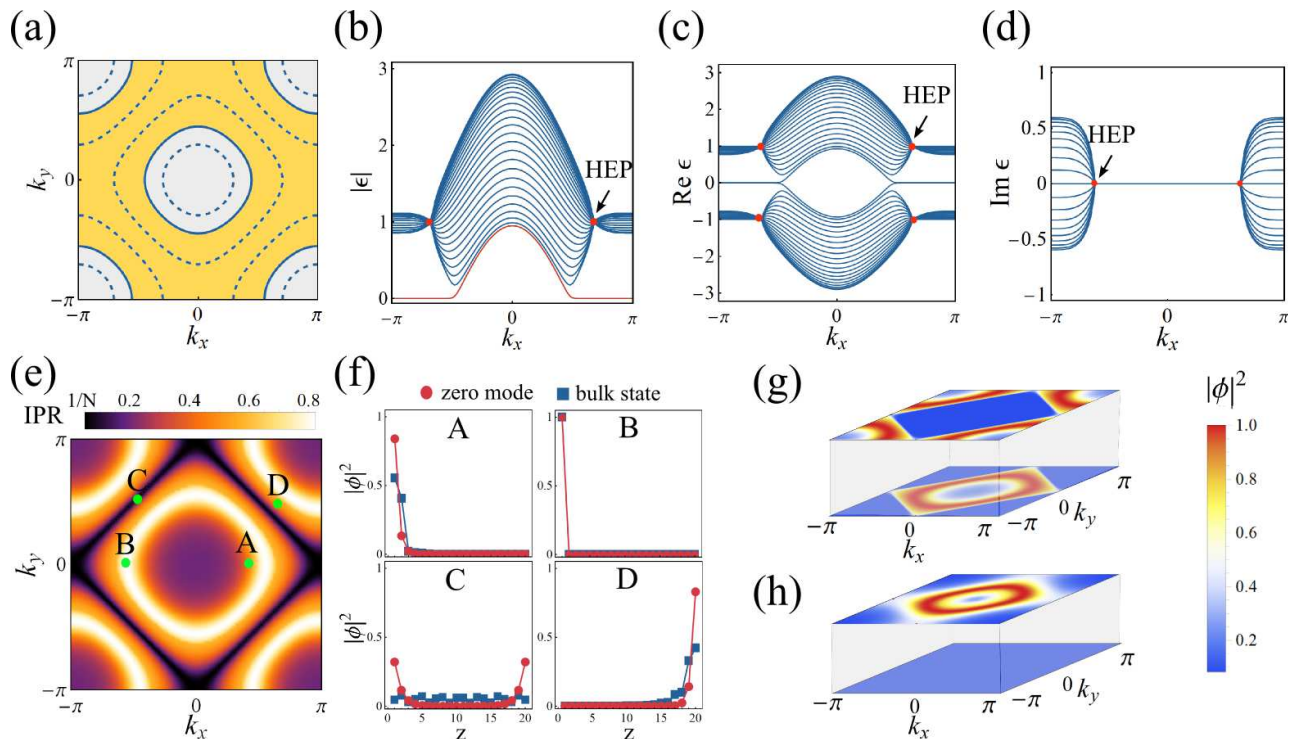}
\caption{(a) Under OBCs in the $z$ direction with $N=20$ slabs, illustration of the zero-energy drumhead-like surface bands in the yellow region bounded by solid lines, where the projections of bulk ERs (dashed lines) are presented for comparison. Here, $m=3$, $B=0.5$, $V_{z}=1$, and $\gamma_{z}=0.6$ are taken. (b) The absolute, (c) real, and (d) imaginary values of the energy bands are shown along the $k_{x}$ direction for the fixed $k_{y}=0$. (e) The inverse participation ratio (IPR) of a typical bulk eigenstate as a function of both $k_{x}$ and $k_{y}$. Extended states exist only in the black regions around $m_{xy}=0$ with IPR$\approx1/N=0.05$, and IPR reaches its maximum value in the white region around $m_{xy}=\pm\gamma_{z}$. (f) The wave function distribution as a function of the slab position for both the zero mode and the bulk eigenstate for points A, B, C, and D in (e) with $m_{xy}=1$, $\gamma_{z}$, $0$, and $-1/2$, respectively. (g) and (h) The wave function distribution on the top and bottom slabs. The states closest to zero energy with $m=3$, $\gamma_{z}=0.6$ in (g) and $m=0.9$, $\gamma_{z}=1.1$ in (h) are taken as the examples.}
\end{figure*}

By taking $k_{i}\rightarrow \sin k_{i}$ and $k_{i}^{2}\rightarrow2(1-\cos k_{i})$ in Eq. (\ref{Hamiltonian}), the lattice model Hamiltonian can be obtained as
\begin{equation}
\label{lattice model}
\begin{split}
H=&\big[m-2B(3-\cos k_{x}-\cos k_{y}-\cos k_{z})\big]\tau_{x}\\
&+(v_{z}\sin k_{z}+i\gamma_{z})\tau_{z},
\end{split}
\end{equation}
where the $\epsilon_{0}(\mathbf{k})$ term has been dropped for simplicity. Band degeneracies are found to occur in the $k_{z}=0$ plane at
\begin{equation}
\label{kz0}
\cos k_{x}+\cos k_{y}=2-\frac{m\pm\gamma_{z}}{2B}
\end{equation}
and in the $k_{z}=\pi$ plane at
\begin{equation}
\label{kzpi}
\cos k_{x}+\cos k_{y}=4-\frac{m\pm\gamma_{z}}{2B}.
\end{equation}

In the absence of the non-Hermitian $i\gamma_{z}\tau_{z}$ term, a nodal loop appears in the $k_{z}=0$ plane when $0<\frac{m}{2B}<4$, and in the $k_{z}=\pi$ plane when $2<\frac{m}{2B}<6$, as illustrated by the red and blue dashed lines, respectively, in Fig. 3(e), with $m=3$, $B=0.5$, $v_{z}=1$. In the presence of a small $i\gamma_{z}\tau_{z}$ term, analogous to the continuum model, each nodal loop will split into two ERs [see the solid lines in Fig. 3(e) with $\gamma_{z}=0.6$]. The energy is also purely imaginary between the two ERs and purely real outside. With increasing $\gamma_{z}$, each inner [outer] ER shrinks towards $(k_{x},k_{y})=(0,0)$ [$(\pi,\pi)$] and vanishes there beyond a critical value of $\gamma_{z}$ determined by Eqs. (\ref{kz0}) and (\ref{kzpi}). For example, if $\gamma_{z}$ is increased to $1.2$ in Fig. 3(e), only one ER persists in both the $k_{z}=0$ and $k_{z}=\pi$ planes, as shown by the red and blue solid lines, respectively, in Fig. 3(f).

Similar to the continuum model, by treating $k_{x}$ and $k_{y}$ as parameters, the bulk band can also be characterized by the winding number in Eq. (\ref{winding number}), where the integral interval of $k_{z}$ should now be replaced by $[-\pi,\pi]$. Depending on the model parameters, the winding number $w$ may take a value of $-1$, $-1/2$, or $0$. Regions of distinct $w$ are bounded by the ERs, as can be seen from Figs. 3(e) and 3(f). The emergence of such values of $w$ also originate from encircling the EPs, as has already been clarified in the continuum model.

To examine the bulk-surface correspondence in the non-Hermitian nodal-line semimetals, as a first step, we choose OBCs in the $z$ direction of $N=20$ slabs with the same parameters as those in Fig. 3(e) to numerically calculate the spectrum as a function of both $k_{x}$ and $k_{y}$. Zero-energy surface states are found in the yellow regions in Fig. 4(a), where the projections of the bulk ERs under PBCs are also provided for comparison (blue dashed lines). The discrepancy between the boundaries of the zero-energy flat bands (blue solid lines) and the projections of bulk ERs is obvious, which indeed reflects the breakdown of the usual bulk-surface correspondence of Hermitian nodal-line semimetals. This discrepancy can be well explained as follows.

By treating $k_{x}$ and $k_{y}$ as parameters, the Hamiltonian in Eq. (\ref{lattice model}) will be effectively reduced to a 1D one:
\begin{equation}
\label{1D lattice model}
H_{xy}=(m_{xy}+2B\cos k_{z})\tau_{x}+(v_{z}\sin k_{z}+i\gamma_{z})\tau_{z},
\end{equation}
where $m_{xy}=m-2B(3-\cos k_{x}-\cos k_{y})$. This Hamiltonian takes the same form as the 1D non-Hermitian lattice model in Refs.~\cite{Lee2016, Martinez2018}. It also bears a very close resemblance to the well-studied non-Hermitian SSH model after taking a basis change $\tau_{z}\rightarrow\tau_{y}$~\cite{Yao2018Edge, Kunst2018, Lieu2018}. For simplicity, we will choose the parameters $B=0.5$ and $V_{z}=1$. Following Refs.~\cite{Yao2018Edge} and \cite{Kunst2018}, under OBCs in the $k_{z}$ direction, it can be shown that topological phase transitions accompanying the (dis)appearance of boundary zero modes take place at $m_{xy}=\pm\sqrt{\gamma_{z}^{2}+1}$ or $\pm\sqrt{\gamma_{z}^{2}-1}$. This is in striking contrast to the periodic case, where the bulk ERs are projected to $m_{xy}=1\pm\gamma_{z}$ and $-1\pm\gamma_{z}$. Under OBCs, the topologically nontrivial region with boundary zero modes corresponds to~\cite{Yao2018Edge, Kunst2018} (see Appendix C for a detailed calculation)
\begin{equation}
\label{phase}
\left\{
  \begin{array}{ll}
    |m_{xy}|<\sqrt{\gamma_{z}^{2}+1}, & \hbox{for $\gamma_{z}<1$;} \\
    \sqrt{\gamma_{z}^{2}-1}<|m_{xy}|<\sqrt{\gamma_{z}^{2}+1}, & \hbox{for $\gamma_{z}>1$.}
  \end{array}
\right.
\end{equation}
This is numerically verified in Fig. 4(a), where surface flat bands are bounded by blue solid lines characterized by $m_{xy}=\pm\sqrt{\gamma_{z}^{2}+1}$ instead of the dashed lines representing bulk ERs. As a further illustration, we plot the absolute [Fig. 4(b)], real [Fig. 4(c)], and imaginary [Fig. 4(d)] values of the full complex energy spectra as a function of $k_{x}$ with fixed $k_{y}=0$. Although the model is non-Hermitian with gain and loss, in some parameter regions, the spectra become purely real, which may result from a $PT$-like symmetry~\cite{Bender1998, Lee2016, Martinez2018, Yao2018Edge}. Moreover, since both the real and imaginary parts of the flat bands equal zero ($|\epsilon|=0$), they should be dynamically stable zero modes.

\subsection{Non-Hermitian skin effect}
We continue to investigate the exotic non-Hermitian skin effect under OBCs in our system. For the 1D Hamiltonian in Eq. (\ref{1D lattice model}), it can be shown that when $m_{xy}<0$ ($m_{xy}>0$), not only the zero modes but also a macroscopic fraction of the bulk eigenstates may be localized near the top (bottom) boundary for a large parameter region~\cite{Martinez2018}. This stems from the parameter $\beta$ describing the behavior of an eigenstate in the $z$ direction as $\phi(z)=\beta^{n}\phi(z_{0})$, with $z_{0}$ denoting the position of the bottom slab. Obviously, $|\beta|<1$ ($|\beta|>1$) corresponds to a state localized near the bottom (top) surface, and $|\beta|=1$ describes an extended state. According to Ref. \cite{Yao2018Edge}, the bulk eigenstates for a long chain require (see Appendix B for details)
\begin{equation}
\label{beta}
|\beta|=\sqrt{\Big|\frac{m_{xy}-\gamma_{z}}{m_{xy}+\gamma_{z}}\Big|},
\end{equation}
leading to $|\beta|>1$ ($|\beta|<1$) for $m_{xy}<0$ ($m_{xy}>0$) and $|\beta|=1$ for $m_{xy}=0$.

To further characterize the localization property, we calculate the inverse participation ratio (IPR) to measure the localization of a state $\phi_{i}$, which is defined as $\sum_{z}|\phi_{i}(z)|^{4}/[\sum_{z}|\phi_{i}(z)|^{2}]^{2}$~\cite{Martinez2018}. For extended states, it should be proportional to $1/N$, where $N$ is the total lattice number in the open boundary direction. Figure 4(e) numerically shows the IPR of a typical bulk eigenstate with the same set of parameters as in Fig. 4(a). It can be seen that the extended states exist only in the vicinity of the lines characterized by $m_{xy}=0$ (black regions), with $|\beta=1|$ as predicted by Eq. (\ref{beta}), while the maximum IPR appears around the lines with $m_{xy}=\pm\gamma_{z}$ (white regions), where $|\beta|\rightarrow0$ or $\infty$, implying completely localized states. As an illustration, we choose four representative points, A $(\pi/2,0)$, B $(-\arccos(-0.4),0)$, C $(-\pi/2,\pi/2)$, and D $(2\pi/3,\pi/2)$, with $m_{xy}=1$, $\gamma_{z}$, $0$, and $-1/2$, respectively, to plot the wave function distributions $|\phi(z)|^{2}$ in the z direction of both the zero mode and a representative bulk eigenstate in Fig. 4(f). For point A (D), both the zero mode and the bulk eigenstate are localized near the bottom (top) slab, while for point B, both eigenstates are indeed totally localized at the bottom slab, which may also be related to the occurrence of higher-order EPs (HEPs), as marked by red points in Figs. 4(b)-4(d)~\cite{Martinez2018}. For point C, the zero mode is distributed equally on both surfaces, while the bulk state has now become extended.

Intriguingly, depending on the parameter $m$, the surface flatbands and a macroscopic fraction of bulk eigenstates may be localized at (i) the bottom surface when $m>5$ ($m_{xy}>0$ is always satisfied), (ii) the top surface when $m<1$ ($m_{xy}<0$ is always satisfied), (iii) both the top and bottom surfaces but at different surface BZ regions when $1<m<5$.  For example, we plot the wave function distribution of the state closest to zero energy on the top and bottom slabs, respectively, for $m=3$, $\gamma_{z}=0.6$ [Figs. 4(g)] and $m=0.9$, $\gamma_{z}=1.1$ [Fig. 4(h)], where distinct localization behaviors between them can clearly be observed from the wave function distribution on opposite boundary slabs.

\section{Discussion and Conclusion}
Experimentally speaking, although it is quite challenging to tune the gain-and-loss term in condensed matter systems, dissipative waveguide systems and ultracold atomic gas may provide a feasible platform to create and engineer such a non-Hermitian perturbation. For example, the gain-and-loss term for the two ``orbitals'' can be effectively realized in ultracold atomic systems by using a resonant optical beam or applying a radio frequency pulse to generate an effective decay for one of the two orbitals~\cite{Xu2017Weyl}. Moreover, there have already been several proposals~\cite{Zhang2016Quantum, Xu2016Dirac} for realizing nodal-line semimetals in ultracold optical lattices, which was recently experimentally observed~\cite{Song2018}. It is also worth mentioning that non-Hermitian Weyl exceptional rings have been experimentally realized in optical waveguide arrays~\cite{Cerjan2018Experimental}. Very recently, electric-circuit realizations of non-Hermitian topological phases were also proposed in Refs. \cite{luo2018nodal, ezawa2018electric}.

In summary, we have theoretically investigated non-Hermitian nodal-line semimetals, where the non-Hermiticity originates from the introduced particle gain-and-loss perturbation. Through dimensional reduction, two different topological numbers have been used to describe the topology of the bulk bands. By comparing the band structures under PBCs and OBCs, the conventional bulk-surface correspondence in nodal-line semimetals was found to fail in the non-Hermitian case. Furthermore, the non-Hermitian skin effect in our system was also discussed based on the knowledge from 1D non-Hermitian models.

\begin{acknowledgments}
This work was supported by the National Natural Science Foundation of China (No.~11674165), the Fok Ying-Tong Education Foundation of China (Grant No.~161006) and the Fundamental Research Funds for the Central Universities (No. 020414380038).
\end{acknowledgments}

\appendix
\section{Derivation of the winding number}
In this section, through the method introduced in Ref. \cite{Yin2018}, we explicitly calculate the winding number defined by Eq. (\ref{winding number}) in the main text for the Hamiltonian:
\begin{equation}
h=h_{z}\tau_{z}+h_{x}\tau_{x},
\end{equation}
with $h_{x}=m-B(k_{\rho}^{2}+k_{z}^{2})$, $h_{z}=v_{z}k_{z}+i\gamma_{z}$, and $\phi=\arctan(h_x/h_z)$. Here, $m$, $B$, $v_{z}$, and $\gamma_{z}$ are set to be positive without loss of generality. As a complex angle, $\phi$ can be decomposed as $\phi=\phi_{\mathrm{R}}+i\phi_{\mathrm{I}}$, where $\phi_{\mathrm{R}}$ and $\phi_{\mathrm{I}}$ denote the real and imaginary parts of $\phi$, respectively. For later reference, the values of $\phi$ for the two limits $k_{z}\rightarrow\pm\infty$ are given as
\begin{equation}
\phi_{k_{z}\rightarrow\pm\infty}=\arctan\Big(\frac{h_{x}}{h_{z}}\Big)_{k_{z}\rightarrow\pm\infty}=\mp\frac{\pi}{2},
\end{equation}
which are purely real.

Through the relation
\begin{equation}
e^{2i\phi}=\frac{\cos\phi+i\sin\phi}{\cos\phi-i\sin\phi}=\frac{1+i\tan\phi}{1-i\tan\phi}=\frac{h_{z}+ih_{x}}{h_{z}-ih_{x}},
\end{equation}
it is obvious that the amplitude and phase parts are related to $\phi_{\mathrm{I}}$ and $\phi_{\mathrm{R}}$, respectively, as
\begin{equation}
e^{-2\phi_{\mathrm{I}}}=\Big|\frac{h_{z}+ih_{x}}{h_{z}-ih_{x}}\Big|,
\end{equation}
and
\begin{equation}
e^{2i\phi_{\mathrm{R}}}=\frac{h_{z}+ih_{x}}{h_{z}-ih_{x}}\bigg/\Big|\frac{h_{z}+ih_{x}}{h_{z}-ih_{x}}\Big|.
\end{equation}
First, since $\phi_{I}$ is found to be a continuous function of $k_{z}$, the imaginary part of the integral in Eq. (\ref{winding number}) is obtained as
\begin{equation}
\frac{1}{2\pi}\int^{\infty}_{-\infty}dk_{z}\partial_{k_{z}}\phi_{\mathrm{I}}=
\frac{\phi_{\mathrm{I}}(k_{z}\rightarrow\infty)-\phi_{\mathrm{I}}(k_{z}\rightarrow-\infty)}{2\pi}=0.
\end{equation}
Now, consider the relation
\begin{equation}
\tan(2\phi_{\mathrm{R}})=\mathrm{Im}\Big(\frac{h_{z}+ih_{x}}{h_{z}-ih_{x}}\Big)\bigg/\mathrm{Re}\Big(\frac{h_{z}+ih_{x}}{h_{z}-ih_{x}}\Big);
\end{equation}
it can be rewritten as \cite{Yin2018}
\begin{equation}
\tan(2\phi_{\mathrm{R}})=\tan(\phi_{A}+\phi_{B}),
\end{equation}
with the two real angles defined via \cite{Yin2018}
\begin{equation}
\begin{split}
\tan\phi_{A}=&\frac{\mathrm{Re}(h_{x})+\mathrm{Im}(h_{z})}{\mathrm{Re}(h_{z})-\mathrm{Im}(h_{x})}=\frac{m-B(k_{\rho}^{2}+k_{z}^{2})+\gamma_{z}}{v_{z}k_{z}}\\
\tan\phi_{B}=&\frac{\mathrm{Re}(h_{x})-\mathrm{Im}(h_{z})}{\mathrm{Re}(h_{z})+\mathrm{Im}(h_{x})}=\frac{m-B(k_{\rho}^{2}+k_{z}^{2})-\gamma_{z}}{v_{z}k_{z}}.
\end{split}
\end{equation}
Then we can simply get
\begin{equation}
\phi_{\mathrm{R}}=n\pi+\frac{1}{2}(\phi_{A}+\phi_{B}),
\end{equation}
where $n$ is an integer. Note that both $\phi_{A}$ and $\phi_{B}$ exhibit discontinuities at $k_{z}=0$, with
\begin{equation}
\begin{split}
\phi_{A}(k_{z}\rightarrow 0^{\pm})=&\pm\frac{\pi}{2}\mathrm{sgn}(m+\gamma_{z}-Bk_{\rho}^{2})\\
\phi_{B}(k_{z}\rightarrow 0^{\pm})=&\pm\frac{\pi}{2}\mathrm{sgn}(m-\gamma_{z}-Bk_{\rho}^{2}).
\end{split}
\end{equation}
Moreover, when $k_{z}\rightarrow\pm\infty$,
\begin{equation}
\begin{split}
\phi_{A}(k_{z}\rightarrow \pm\infty)=\phi_{B}(k_{z}\rightarrow \pm\infty)=\mp\frac{\pi}{2}.
\end{split}
\end{equation}
Finally, we have
\begin{equation}
\begin{split}
w=&\frac{1}{2\pi}\int_{-\infty}^{\infty}dk_{z}\partial_{k_{z}}\phi_{\mathrm{R}}\\
=&\frac{1}{4\pi}\int_{-\infty}^{\infty}dk_{z}\partial_{k_{z}}(\phi_{A}+\phi_{B})\\
=&\frac{1}{4\pi}\bigg(\Big(\phi_{A}\big|^{+\infty}_{0^{+}}+\phi_{A}\big|_{-\infty}^{0^{-}}\Big)
+\Big(\phi_{B}\big|^{+\infty}_{0^{+}}+\phi_{B}\big|_{-\infty}^{0^{-}}\Big)\bigg)\\
=&-\frac{1}{2}-\frac{\mathrm{sgn}(m+\gamma_{z}-Bk_{\rho}^{2})+\mathrm{sgn}(m-\gamma_{z}-Bk_{\rho}^{2})}{4}\\
=&\left\{
   \begin{array}{ll}
     -1, & \hbox{$|k_{\rho}|<\sqrt{m-\gamma_{z}}$;} \\
     -\frac{1}{2}, & \hbox{$\sqrt{m-\gamma_{z}}<|k_{\rho}|<\sqrt{m+\gamma_{z}}$;} \\
     0, & \hbox{$|k_{\rho}|>\sqrt{m+\gamma_{z}}$.}
   \end{array}
 \right.
\end{split}
\end{equation}
This is exactly Eq. (\ref{w result}) in the main text.
\section{Derivation of $\beta$ for bulk states under OBCs}
\begin{figure}[t]
\includegraphics[clip,angle=0.3,width=5cm]{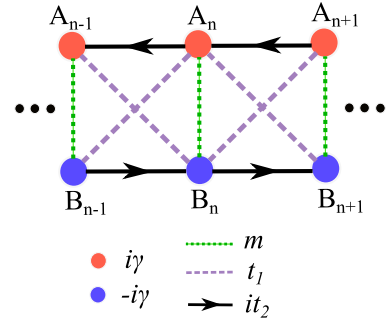}
\caption{Schematic illustration of the 1D tight-binding model for the effective momentum-space Hamiltonian in Eq. (\ref{1D lattice model}) with $\gamma=\gamma_{z}$, $m=m_{xy}$, $t_{1}=B$, and $t_{2}=v_{z}/2$.}
\end{figure}
In this section, we will present a brief derivation of the condition in Eq. (\ref{beta}) for the parameter $\beta$ of bulk states under OBCs. We start from the 1D real-space tight-binding model with two orbitals, $A$ and $B$, in a unit cell for the momentum-space Hamiltonian in Eq. (\ref{1D lattice model})~\cite{Lee2014, Martinez2018}, which is schematically shown in Fig. 5. Here, $t_{1}=B$ represents the intercell interorbital nearest-neighbor (NN) hopping, and $-it_{2}$ and $it_{2}$ with $t_{2}=v_{z}/2$ are the intercell intraorbital NN hoppings for $A$ and $B$ orbitals, respectively, $m=m_{xy}$ denotes the intracell interorbital hopping, and $i\gamma$ ($-i\gamma$) with $\gamma=\gamma_{z}$ is the on-site gain-and-loss term for $A$ ($B$). The real-space wave function satisfies
\begin{widetext}
\begin{equation}
\begin{split}
it_{2}\psi_{An-1}+t_{1}\psi_{Bn-1}+i\gamma\psi_{An}+m\psi_{Bn}-it_{2}\psi_{An+1}+t_{1}\psi_{Bn+1}&=E\psi_{An},\\
t_{1}\psi_{An-1}-it_{2}\psi_{Bn-1}+m\psi_{An}-i\gamma\psi_{Bn}+t_{1}\psi_{An+1}+it_{2}\psi_{Bn+1}&=E\psi_{Bn}.
\end{split}
\end{equation}
\end{widetext}
Analogous to Ref. \cite{Yao2018Edge}, by taking the ansatz $(\psi_{An},\psi_{Bn})=\beta^{n}(\psi_{A},\psi_{B})$, we get
\begin{equation}
\begin{split}
i\Big[t_{2}\big(\frac{1}{\beta}-\beta\big)+\gamma\Big]\psi_{A}+\Big[t_{1}\big(\frac{1}{\beta}+\beta\big)+m\Big]\psi_{B}&=E\psi_{A},\\
-i\Big[t_{2}\big(\frac{1}{\beta}-\beta\big)+\gamma\Big]\psi_{B}+\Big[t_{1}\big(\frac{1}{\beta}+\beta\big)+m\Big]\psi_{A}&=E\psi_{B}.\\
\end{split}
\end{equation}
This leads to the condition
\begin{equation}
\begin{split}
E^{2}+\Big[t_{2}\big(\frac{1}{\beta}-\beta\big)+\gamma\Big]^{2}=\Big[t_{1}\big(\frac{1}{\beta}+\beta\big)+m\Big]^{2},
\end{split}
\end{equation}
from which $\beta$ can be determined. In this paper, we consider the simple case of $t_{1}=t_{2}=t$ ($B=v_{z}/2$), where the above equation can be reduced to
\begin{equation}
\label{betaequation}
\begin{split}
2t(m+\gamma)\beta^{2}+(m^{2}-\gamma^{2}+4t^{2}-E^{2})\beta+2t(m-\gamma)=0,
\end{split}
\end{equation}
leading to two solutions, $\beta_{1}$ and $\beta_{2}$, which satisfy
\begin{equation}
\label{betasolution}
\beta_{1}\beta_{2}=\frac{m-\gamma}{m+\gamma}.
\end{equation}
Through a similar argument in Ref. \cite{Yao2018Edge} for the general solution, it can be shown that the bulk states of a long chain require $|\beta_{1}|=|\beta_{2}|$. Combined with Eq. (\ref{betasolution}), this yields
\begin{equation}
\label{betacondition}
|\beta|=|\beta_{1}|=|\beta_{2}|=\sqrt{\Big|\frac{m-\gamma}{m+\gamma}\Big|},
\end{equation}
which is Eq. (\ref{beta}) in the main text. When $|\beta|<1$ ($|\beta|>1$), the bulk states are localized at the left (right) end, corresponding to the bottom (top) slab in the main text.

\section{Derivation of the topological nontrivial region under OBCs}
Based on Eq. (\ref{betaequation}), in the $E\rightarrow0$ limits, we get
\begin{equation}
\beta_{1,2}=-\frac{m-\gamma}{2t}, \quad-\frac{2t}{m+\gamma}.
\end{equation}
Following Ref. \cite{Yao2018Edge}, the phase boundaries where the bulk states touch zero energy can be determined by inserting Eq. (\ref{betacondition}) into $|\beta_{1,2}|$ from the above equation, leading to
\begin{equation}
m=\pm\sqrt{\gamma^{2}+4t^{2}} \quad\mathrm{or} \quad\pm\sqrt{\gamma^{2}-4t^{2}},
\end{equation}
where $t=B=v_{z}/2=1/2$ is chosen in Eq. (\ref{phase}) of the main text. Then, using the methods introduced in Ref. \cite{Yao2018Edge}, the OBC topological invariant $\chi$ (winding number) for the non-Bloch Hamiltonian obtained by replacing $e^{ik}\rightarrow\beta$ and $e^{-ik}\rightarrow\beta^{-1}$ in Eq. (\ref{1D lattice model}) can be readily calculated as
\begin{equation}
\chi=\left\{
       \begin{array}{ll}
         1, & \hbox{$|m_{xy}|<\sqrt{\gamma_{z}^{2}+1}$,} \\
         0, & \hbox{$|m_{xy}|>\sqrt{\gamma_{z}^{2}+1}$,}
       \end{array}
     \right.
\end{equation}
when $\gamma_{z}<1$ and
\begin{equation}
\chi=\left\{
       \begin{array}{ll}
         1, & \hbox{$\sqrt{\gamma_{z}^{2}-1}<|m_{xy}|<\sqrt{\gamma_{z}^{2}+1}$,} \\
         0, & \hbox{$|m_{xy}|>\sqrt{\gamma_{z}^{2}+1}$ or $|m_{xy}|<\sqrt{\gamma_{z}^{2}-1}$,}
       \end{array}
     \right.
\end{equation}
when $\gamma_{z}>1$, which leads to the topological nontrivial region in Eq. (\ref{phase}).
\bibliography{reference}

\end{document}